\documentclass[aps,prl,twocolumn,showpacs]{revtex4}
\begin{document}





\noindent {\bf Comment on ``Information Capacity of Optical Fiber
Channels with Zero Average Dispersion''} \vspace{3mm}

It is an important problem, both theoretically and practically, to
calculate or estimate the information capacity of a nonlinear
channel. In a recent Letter \cite{Turitsyn03prl} by Turitsyn {\it et
al}., two models of noisy nonlinear fiber channels with zero average
dispersion have been analyzed, and the spectral efficiency $C$ is
estimated to increase asymptotically as
$C\ge\textstyle{\frac{1}{2}}\log_2(S/N)+O(1)$ at high
signal-to-noise ratio $S/N$, in sharp contrast to previous results
that predict the channel capacity to peak and then fall off as
signal power increases
\cite{Mitra01,Tang01a,Tang01b,Narimanov02,Green02opn}. While
previous works relied on approximations, Ref.\cite{Turitsyn03prl}
used elegant field-theoretic formulations and techniques to
analytically calculate the conditional probability density function,
and rigorously derived a lower bound for the asymptotic channel
capacity. However, the essential difference between
Ref.\cite{Turitsyn03prl} and previous works may not be in the
mathematical approaches, rather, it may lie in subtle but crucial
details of the physical models. In particular, the models in
Ref.\cite{Turitsyn03prl} incorporate no bandwidth-limiting mechanism
during transmission or at the end of it, hence are limited in
applicability to practical fiber channels, where such
bandwidth-limiting mechanisms are always present to slow down the
growth of channel capacity as signal power rises.

Although not explicitly stated, both models of
Ref.\cite{Turitsyn03prl} effectively assume spectral conservation of
signal and noise within a band of width $W$, which is not (or yet to
be) justified, and may not correctly reflect the actual physics of
nonlinear channels. On one hand, no mathematical function or
distribution of a signal is known to preserve its spectrum within a
given frequency band when undergoing noisy nonlinear propagation. On
the other hand, it has become an empirical law that nonlinearity in
conjunction with noise always mix signals to generate new frequency
components \cite{Green02opn}. Solitons may be the closest to
spectral preserving among known waveforms. But it is impractical to
establish a single soliton channel utilizing the entire bandwidth of
a fiber transmission line, while wavelength-division-multiplexed
(WDM) solitons interact and mix with noise to generate out-of-band
spectral components. Furthermore, soliton transmissions require
fiber lines with nonzero average dispersion, which escape from the
analytical solution \cite{Turitsyn03prl}. Even frequency- or
phase-modulated signals that might propagate undistorted under Kerr
nonlinearity, the additive noise would cause intensity fluctuations
and eventually spectral broadening. In conclusion, the physical
models in Ref.\cite{Turitsyn03prl} may not be suitable for practical
fiber channels, and the question remains open whether the capacity
could grow unbounded with increasing $S/N$ for noisy, nonlinear, and
bandlimited channels. A more realistic model may include distributed
frequency filtering to the nonlinear Schr\"odinger equation, in
addition to distributive noise. But then the solvability of the
channel and its capacity remain open problems.

Nevertheless, Ref.\cite{Turitsyn03prl} has served greatly as a
reminder to the trivial but often overlooked fact, that the capacity
can never decrease as the {\em permitted but not had to be reached}
bound of signal power increases. Falling off of capacity after
peaking \cite{Mitra01,Tang01a,Tang01b,Narimanov02,Green02opn} simply
indicates inapplicability of either physical models or mathematical
approximations. Ref.\cite{Mitra01} considered a different problem
where WDM channels are not demodulated or processed in a correlated
manner. Ref.\cite{Narimanov02} treated Kerr nonlinearity as
perturbation, which naturally ceases to apply when the signal power
is high. As well demonstrated \cite{Turitsyn03prl}, the use of
Pinsker's formula as done in Refs.\cite{Tang01a} and \cite{Tang01b}
is prone to error. The channel model of Ref.\cite{Tang01a} for a
single-span system assumes nonlinear but dispersionless propagation
followed by noise addition at the end. In particular,
frequency-dependent loss in the transmission fiber is neglected, so
the channel bandwidth is only limited by the transmitter and
receiver. The model admits a trivial argument of unbounded capacity.
The Kerr nonlinearity in such channel becomes oblivious when
constant-intensity modulations such as frequency- or phase-shift
keying are used. Such particular modulations may not be best
suitable to realize the ultimate capacity, but would definitely
establish an asymptotic spectral efficiency of
$constant\times\log_2(S/N)$ at high $S/N$, just as they do for a
normal linear channel \cite{Gibson93}.

\vspace{3mm}

\noindent Haiqing Wei

School of Information Science and Engineering

Lanzhou University, Lanzhou 730000, China, and

oLambda, Inc., San Jose, California 95134

\vspace{3mm}

\noindent David V. Plant

Department of Electrical and Computer Engineering

McGill University, Montreal, Canada H3A-2A7

\vspace{3mm}


\end{document}